\documentclass[twocolumn]{revtex4}
\usepackage[dvips,usenames]{color}
\usepackage{psfig}
\usepackage{amssymb}
\usepackage[english]{babel}

\begin{document}

\title{Pattern of Reaction Diffusion Front in Laminar Flows}

\author{M. Leconte, J. Martin, N. Rakotomalala, D. Salin}
\affiliation{Laboratoire Fluides Automatique et Syst\`emes Thermiques,
Universit\'es P. et M. Curie and Paris Sud, C.N.R.S. (UMR 7608)
B\^atiment 502, Campus Universitaire, 91405 Orsay Cedex, France}

\begin{abstract}
Autocatalytic reaction between reacted and unreacted species may propagate as
solitary waves, namely at a constant front velocity and with a stationary concentration
profile, resulting from a balance between molecular diffusion and chemical reaction.
The effect of advective flow on the autocatalytic reaction between iodate and
arsenous acid in cylindrical tubes and Hele-Shaw cells is analyzed experimentally
and numerically using lattice BGK simulations. We do observe the existence of
solitary waves with concentration profiles exhibiting a cusp and we delineate
the eikonal and mixing regimes recently predicted. 
\end{abstract}

\maketitle

The motion of interfaces and the propagation of fronts resulting
from chemical reactions occur in a number of different areas \cite{scott},
including population dynamics \cite{fisher,kolmogorov} and flame propagation
\cite{zeldo}. It is known that autocatalytic reaction fronts between two reacting
species propagate as solitary waves, namely at a constant front velocity and
with a stationary concentration profile \cite{hanna,martin}. The important
issue of the selection of the front velocity was addressed earlier on, but only
a few cases are well understood, such as the pioneering works of Fisher \cite{fisher}
and Kolmogorov-Petrovskii-Piskunov \cite{kolmogorov} on a reaction-diffusion
equation with second-order kinetics \cite{scott,zeldo,saarloos}. The effect
of advective flow (inviscid and/or turbulent) on reacting systems was analyzed
extensively in the propagation of flames in the context of combustion \cite{zeldo,clavin}.
On the other hand, advective effects on the behavior of autocatalytic fronts
have been only recently addressed \cite{berest,abel,edwards}.
B. F. Edwards \cite{edwards} studied theoretically the effect
of a \( 2D \) laminar flow on an autocatalytic reaction front between two infinite
planes separated by a gap \( b \). In this geometry, the velocity profile is
unidirectional in the direction \( z \) of the flow and is given by Poiseuille's
equation, \( \overrightarrow{U}=U_{M}(1-\zeta ^{2})\overrightarrow{z} \) where
\( U_{M}=1.5\, \, \overline{U} \) is the maximum velocity, \( \overline{U} \)
is the mean velocity, \( \zeta =2x/b \) is the transverse normalized coordinate
and \( \overrightarrow{z} \) is the unit vector parallel to the flow, chosen
as the direction of the front propagation in the absence of flow (see below).
Consider the iodate-arsenous acid reaction described by a third-order autocatalytic
reaction kinetics \cite{scott,hanna,martin}:
\textcolor{black}{\begin{equation}
\label{ADRE}
\frac{\partial C}{\partial t}+\overrightarrow{U}\cdot \overrightarrow{\nabla }C=D_{m}\triangle C+\alpha C^{2}(1-C)
\end{equation}
 where \( C \) is the concentration of the (autocatalytic) reactant iodide,
normalized by the initial concentration of iodate, \( D_{m} \) is the molecular
diffusion coefficient, and \( \alpha  \) is the reaction rate kinetic coefficient.
In the absence of hydrodynamics (\( \overrightarrow{U}=\overrightarrow{0} \)),
Eq.\ref{ADRE} admits a well-known solitary wave solution with front velocity
\( V_{0}=\sqrt{\alpha D_{m}/2} \) and front width \( L_{0}=D_{m}/V_{0} \)
\cite{hanna,martin}. The use of these two quantities to normalize velocities
and lengths in Eq.\ref{ADRE}, leads to two independent parameters \( \eta =b/2L_{0} \)
and \( \varepsilon =\overline{U}/V_{0} \). Reference \cite{edwards} investigated
numerically the solitary wave solution of Eq.\ref{ADRE}, and particularly its
normalized front velocity, \( v=V_{F}/V_{0} \), as a function of \( \varepsilon  \),
for different values of \( \eta  \). Of interest are the following asymptotic
predictions:}

In the narrow-gap regime (\( \eta \rightarrow 0 \) or \( \varepsilon \rightarrow 0 \)),
it was found that \( v=1+\varepsilon  \). Namely, when \( L_{0}>>b \), mixing
across the gap is significant, the concentration front is flat and advected
by the mean flow, yielding: \( V_{F}=V_{0}+\overline{U} \).

On the other hand, in the wide-gap regime (\( \eta \gg 1 \)),
the front is thin and curved across the gap, and Eq.\ref{ADRE} can be replaced
by the eikonal equation: \begin{equation}
\label{eiko}
\overrightarrow{V_{F}}.\overrightarrow{n}=V_{0}+\overrightarrow{U}.\overrightarrow{n}+D_{m}\kappa 
\end{equation}
 where \( \overrightarrow{n} \) is the unit vector normal to the thin front
(oriented from reacted to unreacted species) and \( \kappa  \) the front curvature.
In this regime, to leading order and neglecting the local front curvature, the
front velocity is given by the simplified \( 1D \) eikonal equation: \begin{equation}
\label{eikonal}
V_{F}=V_{0}/\cos \theta +U(\zeta )
\end{equation}
 where \( \theta  \) is the angle between \( \overrightarrow{n} \) and the
flow direction and \( U(\zeta ) \) is the advection velocity. Under these conditions,
reference \cite{edwards} predicted two behaviors depending on the flow direction:
For a supporting flow (\( \varepsilon >0 \)), \( V_{F}=V_{0}+U_{M} \),
which means that the front is advected at the largest possible velocity. The
front shape across the gap is then given by the solution of Eq.\ref{eikonal}.
For an adverse flow (\( \varepsilon <0 \)), \( V_{F}=V_{0} \),
which also represents the maximum algebraic velocity one could have expected
physically. The front is perpendicular to the walls (\( \theta =0 \) at \( \zeta =\pm 1 \)),
and presents a cusp in the middle of the gap (discontinuity of \( \theta  \)
at \( \zeta =0 \)). Here, the adverse flow elongates the front but does not
slow it down. Note that curvature effects (\( D_{m}\kappa  \) in Eq.\ref{eiko})
smooth the cusp, but do not otherwise alter these predictions. Note also that
similar features would occur for other kinetics such as FKPP or Arrhenius ones
\cite{berest}.

\begin{figure}[!h]
\begin{center}
    \begin{minipage}{90mm}
            \psfig{file=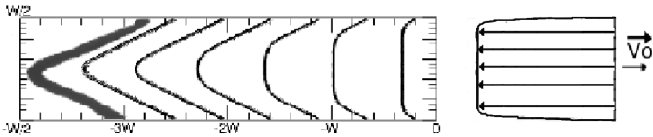,width=90mm,angle=0}
            \caption{\small Time evolution of the front from a flat shape to a triangular 
one in the plane of a Hele-Shaw cell. The fronts obtained with $3D$ lattice BGK
simulations (lines) are compared to the experimental stationary front (left of the figure). 
The figure on the right displays the gap-averaged flow velocity profile of the simulation.}
            \label{num}
    \end{minipage}
\end{center}
\end{figure}

The objective of the present letter is to experimentally test
the above \( 2D \) predictions using two different devices, namely Hele-Shaw
cells and cylindrical tubes. The case of the Hele-Shaw cell, consisting of two
parallel plates separated by a gap \( b \) small compared to the other dimensions,
is supposed to be quantitatively addressed by \cite{edwards}. Alternatively,
the case of the cylindrical tube of inner radius \( a \) (in which the flow
field is also described by Poiseuille's equation, with \( U_{M}=2\overline{U} \)
and \( \zeta =r/a \)) represents a genuine (axisymmetric) \( 2D \) situation.
Experiments in Hele-Shaw cells are discussed with the help of lattice BGK simulations
\cite{nous} of Eq.\ref{ADRE} for a \( 3D \) flow.\\
In the experiment, the front is detected by using starch,
at small concentrations, which reacts in the presence of iodine leading to a
dark blue signature of the front passage \cite{martin,nous}. First, we consider
the reaction in the absence of advection by the flow (\( \overrightarrow{U}=\overrightarrow{0} \)).
As expected, we do observe solitary fronts propagating with flat shapes. In
accordance with \cite{hanna,martin}, their velocity is \( V_{0}\sim 0.02\, \, mm/s \),
from which one can estimate their front thickness \( L_{0}=D_{m}/V_{0}\sim 0.1\, \, mm \)
(\( D_{m}\sim 2.10^{-9}\, \, m^{2}/s \)). Because the reaction products have
a lower density than the unreacted species, the hydrodynamically stable situation
corresponds to descending fronts in vertical tubes. In the following, we focus
on the interplay between advection and propagating fronts. To minimize the effect
of density contrast, we studied the propagation of buoyantly stable fronts in
small cells. We used vertical Hele-Shaw cells of size \( b\times W=0.1\times 1,\, \, 0.2\times 4,\, \, 0.4\times 8,\, \, 1\times 15\, \, mm^{2} \)
and circular capillary tubes of radius \( a=0.3,0.58,0.88 \) and \( 1.9\, \, mm \).
A constant advecting flow, upwards or downwards, was fixed by a syringe. Note
that these cells are small enough to prevent flattening of the front due to
buoyancy, but large enough to enable a constant flow rate injection with our
injection device. The average velocity of the imposed flow ranged between \( 0 \)
and \( \sim 60V_{0} \).\\
In a \( 3D \) Hele-Shaw cell, the flow velocity profile is
unidirectional and depends on the two transverse coordinates, \( x \) and \( y \)
\cite{nicole3}. The profile across the gap is almost parabolic with a gap average
value uniform over the width \( W \), except in a boundary layer of order \( b \),
within which the velocity vanishes (see the gap-average profile on Fig.\ref{num}).
For the three aspect ratios studied, \( W/b=10,15,20 \), we can estimate from
\cite{nicole3}, \( U_{M}/\overline{U}=1.60 \), \( 1.57 \) and \( 1.55 \),
respectively.
\begin{figure}[!h]
\begin{center}
    \begin{minipage}{90mm}
      \psfig{file=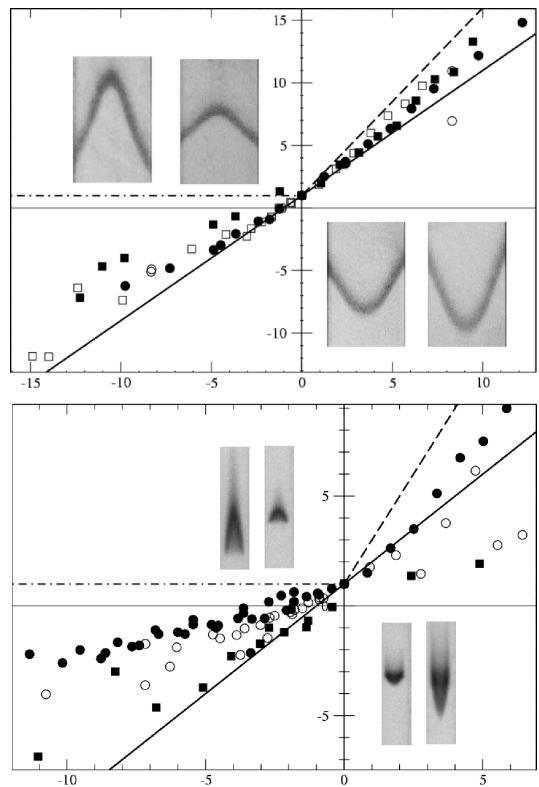,width=70mm,angle=-90}
      \caption{\small Normalized front velocity $v$ versus normalized flow
	velocity $\varepsilon$ 
	 ($\varepsilon < 0$: adverse flow, $\varepsilon > 0$: supportive flow)
for different normalized sizes $\eta$. Top: Hele-Shaw cells of different 
normalized thicknesses and aspect ratios $(\eta=b/2L_0,W/b)$:
$\mbox{\Large$\circ$} (0.5, 10);\mbox{\Large$\bullet$} (1, 20);
\mbox{\scriptsize$\square$} (2,20); \mbox{\scriptsize$\blacksquare$} (5, 15)$.
	Bottom: Circular tubes of different normalized radii ($\eta=a/L_0):
\mbox{\scriptsize$\blacksquare$} (3);\mbox{\Large$\circ$} (5.8);\mbox{\Large$\bullet$} (8.8)$.
	The full and the dashed lines correspond respectively to the mixing regime 
($\eta \rightarrow 0$) and to the eikonal regime ($\eta \rightarrow \infty$).
	Experimental pictures:
	From left to right $\varepsilon=-4.8,-2.4,+2.4,+4.8$ for Hele-Shaw cells
	and $\varepsilon=-6.7,-1.9,+1.9,+6.7$ for tubes.}
      \label{photos}
    \end{minipage}
\end{center}
\end{figure}
We observed solitary waves in the whole range of flow rates
investigated. Typical fronts are shown in Fig.\ref{photos} in the plane of
Hele-Shaw cells (top diagram) and in tubes (bottom diagram). For each geometry,
two adverse flows (on the left) and two supportive flows (on the right) are
displayed. The front shape always points toward the same direction as the underlying
flow field, while its distortion increases with flow intensity (recall that
the fronts are flat in the absence of flow, \( \overline{U}=0 \)). The triangular
shapes observed, in the case of adverse flows, are reminiscent of premixed flames
\cite{mungal}.
\begin{figure}[!h]
\begin{center}
    \begin{minipage}{90mm}
            \psfig{file=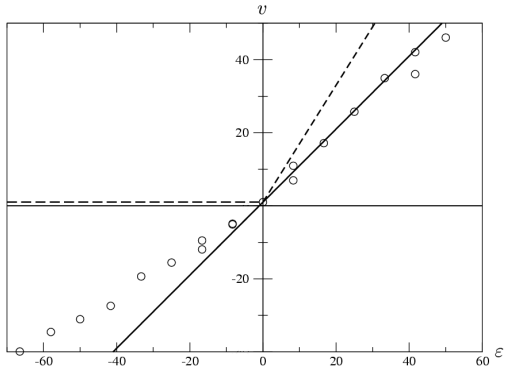,width=90mm,angle=-90}
            \caption{\small Normalized front velocity $v$ versus normalized
	      flow velocity $\varepsilon$ for the smaller Hele-Shaw cell of
	      size $1\times0.1\ mm^2$ ($\eta=0.5, W/b=10$). The full and the dashed lines
	      correspond respectively to the mixing ($\eta \rightarrow 0$) and 
	      to the eikonal ($\eta \rightarrow \infty$) regimes.}
            \label{photos2}
    \end{minipage}
\end{center}
\end{figure}
The two graphs in Fig.\ref{photos} show the normalized front
velocities \( v=V_{F}/V_{0} \) versus \( \varepsilon =\overline{U}/V_{0} \),
measured for different sizes of the Hele-Shaw cells and tubes. For each data
set (given \( \eta  \)), the front velocity increases linearly with the flow
rate, but with a slope different for supportive and adverse flows. This change
of slope is in accordance with \cite{edwards} as well as the observed evolution
of the slopes with \( \eta  \). In addition, most of the data fall in the domain
delimited by the asymptotic regimes described above. The exceptions correspond
to supportive flows in the smaller tubes (\( 1<\eta <6 \)), which, unlike
Edward's \( 2D \) numerical data, fall below the mixing regime (predicted for
\( \eta \ll 1 \)). This difference needs to be further analyzed, given that
it is difficult to achieve experimentally a very low constant flow rate injection
in the small tubes. At the same time, some authors \cite{abel} have suggested
that the mixing straight line should be higher. Using the Peclet number (\( Pe=\overline{U}a/D_{m}=\varepsilon \eta  \)),
which compares the relative importance of advection and diffusion, and the Damk\"{o}hler
number (\( Da=\alpha a/\overline{U}=2\eta /\varepsilon  \)), which is the ratio
of advective to reactive time-scales, they predicted \cite{abel} that for \( Da\ll 1 \)
(which is not attainable in our experiments), the front velocity should be the
product of \( V_{0} \) by the Taylor dispersion factor \cite{Taylor}, which
accounts for the coupling between advection and transverse diffusive mixing.
This factor would then enhance the front velocities.\\
The front velocities measured in the Hele-Shaw cells are very
close to the mixing regime. However, in the case of adverse flows, the measured
values exhibit some departure toward the eikonal regime (\( V_{F}=V_{0} \))
when either \( \eta  \) or \( \varepsilon  \) is increased. This trend is
even more pronounced for the tubes, which present larger \( \eta  \) values
than for the Hele-Shaw cells, in accordance with predictions \cite{edwards}.
On the other hand, all the values measured for supportive flows, even large,
(\( \varepsilon  \) up to \( 50 \) for \( \eta =0.5 \) displayed in Fig.\ref{photos2})
fall on the asymptotic mixing regime predicted by the strictly \( 2D \) gap
analysis \cite{edwards} (\( V_{F}=V_{0}+\bar{U} \) for \( \eta \rightarrow 0 \)).
This is all the more surprising since the flatness of the fronts, expected in
the mixing regime, was not observed in the Hele-Shaw cell plane (top right of
Fig.\ref{photos}). However, in our experiments, although the normalized gap
\( \eta =b/2L_{0} \) introduced in the \( 2D \) gap analysis \cite{edwards}
is small (\( 0.5,1,2,5 \)), the normalized width \( W/2L_{0} \) is large (\( 5,20,40,75 \)).
Extrapolating Edwards's \( 2D \) gap analysis to our \( 3D \) case would suggest
that our experiments combine a mixing regime across the gap with an eikonal
regime across the width. Under these assumptions, the shape and velocity of
the front would obey an equation similar to Eq.\ref{eiko}, namely 
\begin{equation}
\label{eikonal2d}
V_{F}=(V_{0}+D_{m}\kappa )/\cos \theta +U^{2D}(y)
\end{equation}
 where the effective advection velocity \( U^{2D}(y) \) is the gap-averaged
velocity defined in \cite{nicole3}, and where \( \kappa  \) represents now
the curvature of the \( 2D \) front curve observed in the plane of the Hele-Shaw
cell. The front velocity for \( \varepsilon >0 \) would then be set by the
maximum \( U_{M}^{2D} \) of the profile \( U^{2D}(y) \), found in the middle
of the plane (for \( \theta (y=0)=0 \) and \( \kappa (y=0)=0 \)). The so-obtained
maximum front velocity, \( V_{F}=V_{0}+U_{M}^{2D} \), expected in the asymptotic
width-eikonal regime (\( W/2L_{0}>>1 \)) can be compared to the velocity, \( V_{F}=V_{0}+\bar{U} \),
expected in the width-mixing regime (\( W/2L_{0}<<1 \)). One finds that these
two asymptotic velocities would be equal in the Hele-Shaw limit (\( W/b\rightarrow \infty  \)),
and are actually very similar in our experiments (as \( U_{M}^{2D}/\overline{U}=1.07 \),
\( 1.05 \) and \( 1.03 \) for the three aspect ratios used). This could justify
that the parameter \( \eta =b/2L_{0} \) introduced in the \( 2D \) gap analysis
\cite{edwards}, actually controls the front velocity in \( 3D \) Hele-Shaw
cells.
\begin{figure}[!h]
  \begin{center}
    \begin{minipage}{75mm}
      \psfig{file=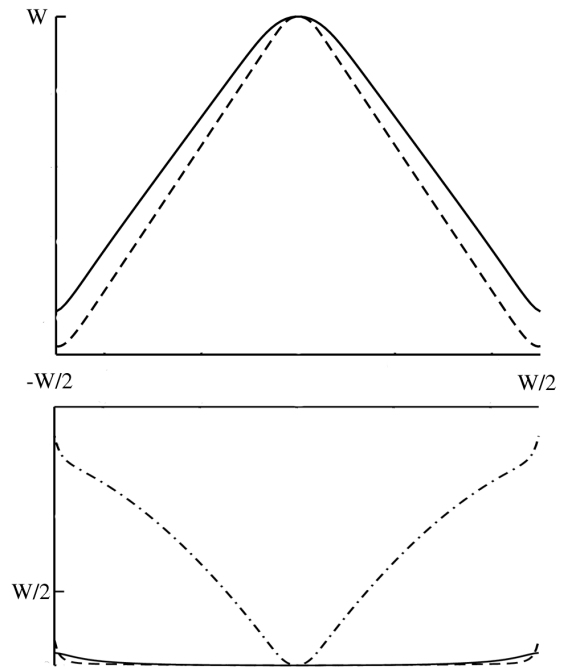,width=75mm,angle=0}
      \caption{\small Calculated concentration fronts in the plane of the cell, obtained by 
$3D$ lattice BGK simulations (full lines) and by integration of the $2D$ eikonal 
Eq.\ref{eikonal2d} (dashed lines), for $\eta=1$ and for one adverse flow 
(top: $\varepsilon=-4.8$) and one supportive flow (bottom: $\varepsilon=4.8$).
The front velocities so-obtained are $V_f/V_0=-3.3$ and $6.16$, respectively.
The dotted-dashed line is the front obtained by the $2D$ eikonal, 
when the flow velocity is slightly modified ($2\%$) to mimic the effect of the meniscus 
which appears on the top of the reacted mixture.}
      \label{eikonum}
    \end{minipage}
  \end{center}
\end{figure}

We tested the ability of the full description and the simplified
one, given respectively by Eqs. \ref{ADRE} and \ref{eikonal2d}, to account
for both shape and velocity of the experimental fronts. As lattice BGK simulations
have been used to obtain the solutions of Eq.\ref{ADRE}, we have first validated
this numerical method, by reproducing Edwards's results \cite{edwards} on the
shape and velocity of the fronts propagating between two infinite planes (\( 2D \)
simulations). Then, \( 3D \) lattice BGK simulations of Eq.\ref{ADRE} and
numerical integration of the \( 2D \) eikonal Eq.\ref{eikonal2d} were performed,
using respectively the analytical stationary \( 3D \) flow field given by \cite{nicole3},
and its gap-average \( U^{2D}(y) \). The front obtained with the latter method
was compared to the iso-concentration \( C=0.5 \) of the gap-averaged concentration
map produced by the lattice BGK simulations. Fig.\ref{eikonum} displays these
fronts, for the same parameters (\( \varepsilon  \), \( \eta  \), \( W/L_{0} \))
as in one typical experiment, in the cases of adverse (\( \varepsilon =-4.8 \))
and supportive (\( \varepsilon =+4.8 \)) flows. Note that in these cases of
interest, for which \( W/L_{0} \) is finite, the integration of Eq.\ref{eikonal2d}
requires the value of the front velocity (thus fixing the value of the curvature
\( \kappa  \) at the integration starting point). Hence, Eq.\ref{eikonal2d}
is not fully predictive, but links the shape of the front to its velocity. The
shape and velocity predicted by lattice BGK simulations and the ones given
by Eq.\ref{eikonal2d} are found to compare fairly well with the experimental
observations in the case of adverse flow. However, for the supportive flow case,
although the front velocity is correctly predicted, the two numerical predictions,
similar for the shape, fail to account for the experimental observations. We
believe that this discrepancy might be due to an alteration of the flow velocity
profile caused by a triangular meniscus which appears on the top of the solution
in our supportive flow experiments. From its shape and its distance to the front
(typically several tens of \( W \)), one can infer that the meniscus could
introduce a few percents of excess fluid velocity in the middle of the cell
plane. The resulting non-uniformity in \( U^{2D}(y) \) may account for the
rounded shape observed in the supportive flow experiments (see Fig.\ref{num}).

We have also analyzed the dynamics of the shape formation
in the case of adverse flows. Fig.\ref{num} displays the time development of
the iso-concentration \( C=0.5 \), initially flat, toward the stationary triangular
shape. The sequence shows that an early determination of both the final front
velocity and the final angle \( \theta  \) is achieved as soon as the profile
is altered over a typical distance \( b \) from the side walls. This supports
the contention that \( W \) plays no role in the determination of both shape
(\( \theta  \)) and velocity in the regimes under consideration. This was confirmed
by simulations in wider lattices which produced the same values of the velocity
and \( \theta  \). Thus \( \eta =b/2L_{0} \) is effectively the relevant parameter
in Hele-Shaw cells (for which \( W/b\gg 1 \)).\\
In conclusion, we have performed experiments and lattice BGK
simulations of autocatalytic reaction fronts in laminar advective flow fields
in Hele-Shaw cells and circular tubes. Solitary waves were observed in the entire
range of flow rates. For flows adverse to the chemical front propagation, we
observed cusp-like fronts in tubes and triangular fronts in the plane of Hele-Shaw
cells. Our measurements of the front velocity agree with the \( 2D \) asymptotic
predictions \cite{edwards}, in the limiting cases where either diffusion overcomes
reaction (\( \eta \ll 1 \)) or it is negligible (\( \eta \gg 1 \)). It would
be interesting to extend the range of the cell sizes. Larger cells could be
used to study the buoyancy stabilizing effect and smaller cells within the scope
of microfluidics.

\textit{This paper benefited from discussions with Professor K. Showalter. The
work was partly supported by IDRIS (project 024052), CNES No 793/CNES/00/8368,
ESA (No AO-99-083). MRT grant (M. L.). All these sources of support
are gratefully acknowledged.}

\end{document}